# High-temperature stable refractory high-entropy nanoalloys with enhanced sinterability

Mingde Qin, Sashank Shivakumar, Jian Luo[*]

Department of NanoEngineering; Program of Materials Science and Engineering,
University of California San Diego, La Jolla, CA, 92093, USA

**ABSTRACT**

Nanocrystalline alloys (nanoalloys) are prone to grain growth. It is known that grain boundary segregation and precipitation can stabilize nanoalloys, but the stabilization becomes less effective at high temperatures and adding grain growth inhibitors often reduces sinterability. Herein, we have simultaneously achieved improved sinterability and exceptional high-temperature stability for a class of MoNbTaTiW-based refractory high-entropy nanoalloys (RHENs). Bulk pellets of RHENs were fabricated through planetary ball milling and spark plasma sintering, achieving 93-96% relative densities with 50-100 nm grain sizes for three compositions. For example, $Mo_{17.8}Nb_{17.8}Ta_{17.8}Ti_{17.8}W_{17.8}Ni_6Zr_5$ sintered at 1300 °C attained ~96% relative density with ~55 nm mean grain size. Moreover, these RHENs exhibited exceptional stability at 1300 °C. Both $Ti_{17.8}Nb_{17.8}Mo_{17.8}Ta_{17.8}W_{17.8}Ni_6Zr_5$ and $Mo_{18.8}Nb_{18.8}Ta_{18.8}Ti_{18.8}W_{18.8}Ni_6$ retained <150 nm grain sizes with >96% of the theoretical densities after five hours annealing at 1300 °C. Notably, the addition of Ni, a well-known sintering aid for activated sintering of refractory metals such as W and Mo, in high-entropy MoNbTaTiW can promote sintering while maintaining high-temperature stability against rapid grain growth, which can be explained by hypothesized effects of high-entropy grain boundaries. These RHENs possess some of the highest temperature stability achieved for nanoalloys and ultrafine-grained metals.

---

[*] Corresponding Author. Email: jluo@alum.mit.edu (Jian Luo)

## I. INTRODUCTION

Nanocrystalline metals show great potential as structural materials through improvements in strength, hardness, and wear resistance that stem from a reduction in grain size.[1-6] However, their applications are limited due to poor thermal stability.[4-6] Specifically, a large amount of grain boundaries (GBs) lead to a considerable increase in the total interfacial (GB) energy, thereby causing rapid and uncontrolled grain growth – in some cases, even at room temperature.[4-7] The stabilization of nanocrystalline alloys (nanoalloys) is commonly achieved through kinetic (impurity solute-drag effects[8] and/or secondary particle Zener pinning[9]) and/or thermodynamic (segregation induced GB energy reduction[10-18]) effects. Using these strategies, stable binary and ternary nanoalloys have been reported.[4,5,10-18] However, both thermodynamic and kinetic stabilization mechanisms are significantly weakened at high temperatures due to thermally induced GB desorption (de-segregation), coarsening of pinning particles, and increased GB mobilities, resulting in loss of desirable properties.[4,5,19-22] Nanocrystalline high-entropy alloys (HEAs) have also been made, which appear to exhibit enhanced stability against grain growth in comparison with their lower-entropy conuterparts.[23-25] In 2016, Zhou *et al* introduced the concept of high-entropy grain boundaries (HEGBs)[26,27] and experimentally demonstrated the stabilization of nanoalloys at elevated temperatures using HEGBs.[27] The theory of HEGBs has been further elaborated recently in a Perspective article.[28] Of particular interest to this study is the so-called "Type II HEGBs" (*i.e.*, HEGBs in HEAs), where bulk high-entropy effects can be used to stabilize nanocrystalline HEAs with one strong segregating element at high temperatures.[27,28] Specifically, it was conjectured[27,28] that adding a small amount of one segregating element in HEA grains can provide improved high-temperature stability against grain growth (in comparison with lower-entropy alloys) via both thermodynamic and kinetic effects, a hypothesis that motivated this study.

Furthermore, adding grain growth inhibitors can often reduce the sinterability (if the nanoalloys are fabricated via powder metallurgy routes). Here, it is well known that certain sintering aids can be used to enhance sintering (*e.g.*, promoting solid-state activated sintering via the formation of liquid-like GBs with high mass transport rates), which can also promote grain growth because of the enhanced GB kinetics.[29-34] Here, suppressing grain growth via reducing GB motion and enhancing sintering via increasing GB diffusion often represent two contradictory requirements. In this study, we found, somewhat surprisingly, that the addition of Ni, a well-known sintering aid for enabling solid-state activated sintering of W and Mo,[29-34] in high-entropy

MoNbTaTiW nanoalloys can promote sintering without accelerating grain growth, which can be explained from the hypothesized HEGB effects.

Motivated by prior studies of activated sintering (*i.e.*, adding a sintering aid to promote densification via segregation-enhanced GB diffusion)[29-34] and the hypothesized theory of Type II HEGBs (*i.e.*, adding a segregating element in nanocrystalline HEAs to inhibit grain growth at high temperatures)[27,28], we added Ni (a well-known sintering aid for W and Ni via forming liquid-like GBs)[29-34] and/or Zr (a more refractory segregant in refractory metals) in NbMoTaTiW HEAs to test their effects on sintering and grain growth. Notably, we have demonstrated simultaneous improvements in sinterability and high-temperature stability against rapid grain growth to attain high relative densities of 93-96% and small mean grain sizes of 50-100 nm for three refractory high-entropy nanoalloys (RHENs). For example, a sintered $Mo_{17.8}Nb_{17.8}Ta_{17.8}Ti_{17.8}W_{17.8}Ni_6Zr_5$ achieved ~96% relative density with a small mean grain size of ~55 nm. Moreover, both $Ti_{18.8}Nb_{18.8}Mo_{18.8}Ta_{18.8}W_{18.8}Ni_6$ and $Ti_{17.8}Nb_{17.8}Mo_{17.8}Ta_{17.8}W_{17.8}Ni_6Zr_5$ retained <150 nm grain sizes with >96% of the theoretical densities after annealing at 1300 °C for 5 hours. The 1300 °C stability represents some of the highest temperature stability that has been achieved for nanoalloys and ultrafine-grained metals.

## II. EXPERIMENTAL

RHENs were synthesized by mixing elemental powders of Mo, Nb, Ta, Ti, W, Ni, and Zr (> 99.5% purity, 325 mesh, Alfa Aesar, MA, USA). For each composition, appropriate amounts of powders were weighed out in batches of 20 g and planetary ball milled (using a mill from Across International, NJ, USA) for 24 h at 300 RPM. Isopropyl Alcohol (IPA) was used as a processing agent, and polytetrafluoroethylene (PTFE) vials and $Y_2O_3$-stablized $ZrO_2$ (YSZ) grinding media (at a ball-to-powder ratio of 10:1) were used for the milling process. After milling, the vials were transferred into a vacuum furnace and dried overnight at 75 °C to remove the processing agent, and then immediately transferred into a low oxygen environment (Ar glovebox, < 15 ppm $O_2$) to prevent oxidation of the fine powders. The powders were then loaded into 10 mm graphite dies (Cal Nano, CA, USA) lined with graphite and Mo foils in batches of 3 g and subsequently consolidated into dense pellets via spark plasma sintering (SPS) in vacuum ($10^{-3}$ - $10^{-2}$ torr) using a Thermal Technologies 3000 series SPS machine. A ramp rate of 100 °C/min was utilized and a hold time of 5 min was employed under 50 MPa of pressure at isothermal sintering temperatures.

Samples were furnace cooled under vacuum. In total, specimens of 26 different compositions and SPS conditions were sintered, which are listed in Supplementary Table S1.

Isothermal annealing was performed on seven series of specimens at 1100 °C, 1200 °C, and 1300 °C, respectively, for 5 hours in a tube furnace under flowing Ar + 5% $H_2$ forming gas. In total, 21 specimens of seven compositions were annealed at three different temperatures, which are listed in Supplementary Table S1, along with the 7 as-sintered specimens as references.

After sintering, bulk specimens were ground to remove Mo foils as well as the carbon-contaminated surface layers caused by the graphite tooling. XRD was performed on a Rigaku Miniflex diffractometers (Cu K$\alpha$ radiation, 30 kV and 15 mA). Density measurements were taken using the Archimedes method. Theoretical densities were calculated from ideal stoichiometry and the lattice parameters measured by X-ray diffraction (XRD). Rockwell hardness (HRC) measurements were taken using a Wilson 574 series Rockwell tester (Buehler, IL, USA) equipped with a diamond spheroconical indenter abiding by the ASTM standard E18-15 with over 30 measurements to ensure the validity.

Samples were fractured after embrittling in liquid $N_2$ and scanning electron microscopy (SEM) was subsequently performed using a Thermo-Fisher Apreo microscope to examine the fractured surfaces. The grain size distributions were obtained from the corresponding SEM images following the ASTM standard E112-88.

## III. RESULTS AND DISCUSSION

### A. Design of Compositions and Overview of Sintering and Grain Growth Experiments

MoNbTaTiW or $Mo_{0.2}Nb_{0.2}Ta_{0.2}Ti_{0.2}W_{0.2}$ is our baseline composition that is denoted as "RHEH-0". This composition was then doped with 6% Ni ($Ti_{18.8}Nb_{18.8}Mo_{18.8}Ta_{18.8}W_{18.8}Ni_6$, denoted as "RHEN-6Ni") and 5% Zr ($Ti_{19}Nb_{19}Mo_{19}Ta_{19}W_{19}Zr_5$, denoted as "RHEN-5Zr"). Furthermore, we tested two co-doped compositions, $Ti_{17.8}Nb_{17.8}Mo_{17.8}Ta_{17.8}W_{17.8}Ni_6Zr_5$ (denoted as "RHEN-6Ni-5Zr") and $Ti_{18.9}Nb_{18.9}Mo_{18.9}Ta_{18.9}W_{18.9}Ni_3Zr_{2.5}$ (denoted as "RHEN-3Ni-2.5Zr").

Specimens of these five different compositions were sintered by SPS at different temperatures (at 1200 °C, 1300 °C, 1400 °C, and 1500 °C, respectively, for all five compositions and at 1100 °C, 1600 °C, and 1700 °C for selected compositions). For example, RHEN-6Ni sintered at 1300

°C by SPS is denoted as "RHEN-6Ni-1300SPS". The sintered bulk specimens were all pellets with a fixed diameter of 10 mm and thicknesses of ~3 mm. Measured relative density and grain size *vs.* SPS temperature curves for all five series of sintered specimens are shown in Fig. 1. Both relative density and grain size of each composition increased with SPS temperature (Fig. 1), which is well expected. Fig. 2 further plots grain size *vs.* relative density curves for five series of specimens after SPS, where the mean grain size generally increased with increasing relative density, following similar trends (but with shifts in the curve positions). In comparison with RHEN-0, Ni-doping and (Ni + Zr) co-doping shifted the grain size *vs.* relative density curves towards right (*i.e.*, higher density with smaller grain sizes, which is desirable), while Zr-doping shifted the curve towards left (undesirable). The sintered specimens consisted primarily of BCC phases, with minor amounts of (expected) secondary phases. For example, XRD patterns for RHEN-6Ni and RHEN-6Ni-5Zr synthesized at 1300 °C showed primary BCC phases (Supplementary Fig. S3); however, minor peak splitting suggested some compositional inhomogeneity after SPS (not surprising given the short SPS time of only 5 min) and additional low-intensity peaks corresponding to minor secondary phases were also observed. The presence of minor secondary phases was well expected since we doped Ni and Zr beyond the solid solubility limits (Table 3) and some native oxides and other impurities/contaminations inevitably existed and could also form during ball milling. Note that we specifically designed our doping levels to above the five *X*-Ni and two *X*-Zr binary solubility limits so we expected multiple secondary precipitation phases (*e.g.*, equilibrium binary $Mo_7Ni_7$, $Ta_2Ni$, $Nb_7Ni_6$, $W_2Zr$, and $Mo_2Zr$ compounds as based on the binary phase diagrams shown in Supplementary Figs. S9-S10, as well as ternary and multicomponent compounds at 1300 °C, along with additional precipitates during cooling) to be present, which are difficult to fully identified due to the large number (up to five or more based on the Gibbs phase rule) of the secondary phases and the small amounts of each phase. Such secondary phases can provide some Zener pinning (regardless the specific phases). Selected as-sintered RHENs that achieved 93-96% relative densities with 50-100 nm grain sizes are listed in Table 1. Notably, RHEN-6Ni-5Zr-1300SPS ($Mo_{17.8}Nb_{17.8}Ta_{17.8}Ti_{17.8}W_{17.8}Ni_6Zr_5$, SPS at 1300 °C) attained ~96% relative density with ~55 nm grain size. Results of all 26 as-sintered specimens made in this study are documented in Supplementary Table S1. These sintering results, particularly the effects of adding Ni and/or Zr, will be further discussed in the subsequent sections.

Grain growth experiments were performed on seven series of selected as-sintered RHENs by annealing them isothermally at 1100 °C, 1200 °C, and 1300 °C, respectively, for 5 hours. Table S2 documents a detailed account of SPS temperature, annealing conditions, densities, and measured grain sizes of all 21 annealed specimens, along with 7 as-sintered specimens before annealing as the references. Fig. 3(a) and (b) plot the changes in relative density and grain size with annealing temperature. All as-sintered compositions achieved relative densities > 85%, with grain sizes < 100 nm, except that the base RHEN-0-1600SPS exhibited the largest as-sintered grain size of ~105 nm. Grain size and relative densities increased with annealing temperatures as expected, and all compositions retained average grain sizes < 150 nm even after 5 h of annealing at 1300 °C. These grain growth results will also be further discussed in the subsequent sections.

We also measured Rockwell hardness for three selected RHENs (RHEN-6Ni-1300SPS, RHEN-6Ni-1400SPS, and RHEN-6Ni-5Ni-1300SPS) after annealing at 1300 °C for 5 h, all of which exhibited similar (96.3-97.2%) relative densities and (124-133 nm) grain sizes (Table 2). The measured hardness values were 64 ± 3 HRC (RHEN-6Ni-1300SPS), 65 ± 2 HRC (RHEN-6Ni-1400SPS), and 62 ± 4 HRC (RHEN-6Ni-5Zr-1300SPS), respectively, as shown in Table 2.

## B. The Benchmark RHEN-0 ($Mo_{0.2}Nb_{0.2}Ta_{0.2}Ti_{0.2}W_{0.2}$)

The benchmark RHEN-0 ($Mo_{0.2}Nb_{0.2}Ta_{0.2}Ti_{0.2}W_{0.2}$) specimens generally showed limited sinterability with small grain sizes. The relative density of as-sintered RHEN-0 increased from ~72.1% (with a small mean grain size of ~11 nm) at a low SPS temperature of 1200 °C to 90.7% (with the mean grain size being increased by almost 10-fold to ~105 nm) at a high SPS temperature of 1600 °C (Fig. 1 and Supplementary Table S1). We further annealed the densest RHEN-0-1600SPS (SPS at 1600 °C) isothermally at 1100 °C, 1200 °C, and 1300 °C, respectively, for 5 hours, which resulted in limited further densification and grain growth (Fig. 2 and Supplementary Table S2). For example, annealing at 1300 °C for 5 hours increased the relative density of RHEN-0 moderately from ~90.7% to ~91.2%, with a moderate increase of the mean grain size from ~105 nm to ~127 nm.

## C. The Effects of Zr *vs.* Ni Doping

Here, we selected Ni and Zr as two doping (alloying) elements, both of which segregate at GBs of *X* = Mo, Nb, Ta, Ti, or W in all ten *X*-Ni and *X*-Zr binary alloys (Table 3). The binary segregation enthalpies estimated using a Miedema-based model by Murdoch and Schuh[35] range from -24,897 J/mol for Ti-Ni to -78,276 J/mol for W-Ni, with a rule-of-mixture (RoM) average of -54,486 J/mol for the five *X*-Ni alloys, and range from -1,280 J/mol for Ti-Zr to -58,933 J/mol for W-Zr, with a RoM average of -31,552 J/mol for the five *X*-Zr alloys (Table 3). Except for Ti-Zr, all other nine binary alloys can be considered as strong segregation systems. Here, it should be noted that we adopt a sign convention where a negative $\Delta H_{seg}$ value suggests GB enrichment or positive segregation[36] (differing from the sign convention used in the original Ref. 35 where a positive $H_{seg}$ infers GB segregation). While both Ni and Zr are strong segregating elements in refractory metals, Ni is a well-known sintering aid for enabling solid-state activated sintering of W, Mo, and other refractory metals (via forming liquid-like interfacial phases),[29-34] but Zr is not. This difference is due to that Zr is a more refractory alloying element than Ni (so it is less likely to promote sintering via forming liquid-like interfacial phases), which is consistent with our observations that the Ni addition promoted sintering in RHEA-6Ni, while the Zr addition suppressed sintering in RHEA-5Zr (Fig. 1a). The melting temperature of Ni ($T_m^{Ni}$ = 1455 °C) is 400 °C lower that of Zr ($T_m^{Zr}$ = 1855 °C). The lowest eutectic temperatures in the five *X*-Ni binary alloys are >200 °C lower than those of the corresponding *X*-Zr binary alloys (Table 1). Notably, the average solidus temperature for the five *X*-Ni binary alloys is 1608 °C, while that for the five *X*-Zr binary alloys is 2021 °C (Table 3). It is noted that we intentionally selected the Ni doping level (6%) that are higher than the solid solubility limits of all five *X*-Ni binary systems (Table 3); thus, we expect the formation of up to five secondary precipitation phases (according to the Gibbs phase rule) but all in small amounts. This can enable us to realize the hypothesized bulk high-entropy stabilization effects via Type II HEGBs, which will be discussed in Section E.

On the one hand, the addition of 5% Zr slightly reduced both densification and grain growth, as shown by the light blue (RHEN-0) *vs.* orange (RHEN-5Zr) curves in Fig. 1 and Fig. 2. On the other hand, the addition of 6% Ni significantly increased both densification and grain growth, as shown by the light blue (RHEN-0) *vs.* green (RHEN-6Ni) curves in Fig. 1 and Fig. 2. For example, at SPS temperature of 1300 °C, addition of 6% Ni increased the relative density from ~83.7% to ~93.1% and ~93.6%, respectively (for two independent specimens), and increased the mean grain size from ~28 nm to ~74 nm and ~78 nm, respectively. Likewise, at SPS temperature of 1400 °C, addition of 6% Ni increased the relative density from ~85.3% to ~94.9% and ~94.7, respectively (for two independent specimens), and increased the mean grain size from ~56 nm to ~103 nm and ~109 nm, respectively. In each case, two RHEN-6Ni specimens were prepared and sintered at normally identical conditions, showing good repeatability of the densification and grain growth results (Fig. 5a,b and Supplementary Fig. S3 and S4).

We annealed the densest RHEN-5Zr-1600SPS. Similar to the undoped RHEN-0-1600SPS, isothermal annealing of RHEN-5Zr-1600SPS at 1100-1300 °C resulted in limited further densification and grain growth (Fig. 2 and Supplementary Table S2). For example, annealing at 1300 °C for 5 hours increased the relative density of RHEN-5Zr moderately from ~87.4% to ~89.7%, with a moderate increase in the mean grain size from ~92 nm to ~117 nm.

We further annealed two series of Ni-doped specimens, RHEN-6Ni-1300SPS and RHEN-6Ni-1400SPS, isothermally at 1100-1300 °C for 5 hours, which resulted in some further densification with limited grain growth (Fig. 2 and Supplementary Table S2). For example, annealing of RHEN-6Ni-1300SPS at 1300 °C for 5 hours increased the relative density of RHEN-6Ni-1300SPS from ~93.1% to ~97.0%, with a moderate increase in the mean grain size from ~74 nm to ~133 nm, as shown in Fig. 5c. In addition, annealing of RHEN-6Ni-1400SPS (that was sintered at a higher SPS temperature of 1400 °C for 5 min to higher density and grian size) at 1300 °C for 5 hours increased the relative density of RHEN-6Ni-1400SPS from ~94.9% to ~97.2%, with a smaller increase of the mean grain size from ~103 nm to ~124 nm, as shown in Fig. 5d. Interestingly, the final densities and grain sizes of the two series of specimens after isothermal annealing are similar, regardless the initial SPS temperatures (1300 °C *vs.* 1400 °C). It appears that the isothermal annealing reduced the initial differences in these two series of RHEN-6Ni specimens (Fig. 4 and Supplementary Table S2).

Notably, both $Mo_{18.8}Nb_{18.8}Ta_{18.8}Ti_{18.8}W_{18.8}Ni_6$ (RHEN-6Ni-1300SPS and RHEN-6Ni-1400SPS) specimens retained <150 nm grain sizes with >96% of the theoretical densities after five hours annealing at 1300 °C, representing some of the best high-temperature stability achieved.

**D. Ni and Zr Co-Doped RHENs**

We further examined two Ni and Zr co-doped compositions, RHEN-6Ni-5Zr (doped with 6% Ni and 5% Zr) and RHEN-3Ni-2.5Zr (doped with 3% Ni and 2.5% Zr), both of which showed enhanced sintering in comparison with RHEN-0 (as shown by the purple/red *vs.* light blue curves in Fig. 1a and Fig. 2).

In particular, RHEN-6Ni-5Zr showed further enhanced sintering in comparison with RHEN-6Ni, particularly at low sintering temperatures (as shown by the purple *vs.* light blue curves in Fig. 1a). For RHEN-6Ni-5Zr, the grain growth was limited at SPS temperatures of 1100-1300 °C, but more substantial at SPS temperatures of 1400 °C and 1500 °C (Fig. 1b). The corresponding grain size *vs.* relative density curves for RHEN-6Ni-5Zr *vs.* RHEN-6Ni curves suggested that Zr co-doping enhanced densification via suppressing grain growth (as the curve for RHEN-6Ni-5Zr shifted towards right in Fig. 2).

As an example, Fig. 6 shows the grain size distribution and SEM micrograph (with a view of a larger region in Supplementary Fig. S7) of RHEN-6Ni-5Zr sintered at 1300 °C, where we achieved a high relative density of ~96% and a small mean grain size of ~55 nm. This represents our best as-sintered nanocrystalline specimen with high density and small nanoscale grain size.

In addition, RHEN-3Ni-2.5Zr showed slightly less densification with similar grain growth in comparison with RHEN-6Ni (as shown by the red *vs.* green curves in Fig. 1), for which the corresponding grain size *vs.* relative density curves almost overlap one another (Fig. 2).

We annealed three selected series of Ni and Zr co-doped specimens, RHEN-6Ni-5Zr-1200SPS, RHEN-6Ni-5Zr-1300SPS, and RHEN-3Ni-2.5Zr-1400SPS (Fig. 2 and Supplementary Table S2). Both RHEN-6Ni-5Zr-1300SPS and RHEN-3Ni-2.5Zr-1400SPS exhibited moderate grain growth with virtually no further densification after isothermal annealing at 1100-1300 °C for 5 hours. RHEN-6Ni-5Zr-1200SPS, which started with a lower relative density of ~92.4% and a small grain size of ~29 nm, exhibited more densification and grain growth, reaching a relative density of ~96.1% and a mean grain size of ~146 nm after annealing at 1300 °C for 5 hours.

Fig. 7 shows the grain size distribution and SEM micrograph (with a view of a larger region in Supplementary Fig. S8) of the RHEN-6Ni-5Zr-1300SPS specimen after annealing at 1300 °C for 5 hours, showing ~96.3% relative density and a mean grain size of ~133 nm. Similarly, RHEN-6Ni-5Zr-1200SPS exhibited ~96.1% relative density and ~146 nm grain size after annealing at 1300 °C for 5 hours. Akin to RHEN-6Ni, an interesting observation is that the final densities (~96.1% *vs.* ~96.3%) and grain sizes (~146 nm *vs.* ~133 nm) are similar after annealing at 1300 °C for 5 hours, regardless the initial SPS temperatures (1200 °C *vs.* 1300 °C) for these two RHEN-6Ni-5Zr specimens, where the isothermal annealing reduced the differences in initial relative densities (~92.4% *vs.* ~96.0%) and grain sizes (~39 nm *vs.* ~55 nm).

The two $Ti_{17.8}Nb_{17.8}Mo_{17.8}Ta_{17.8}W_{17.8}Ni_6Zr_5$ specimens (RHEN-6Ni-5Zr-1200SPS and RHEN-6Ni-5Zr-1300SPS), in addition to two $Mo_{18.8}Nb_{18.8}Ta_{18.8}Ti_{18.8}W_{18.8}Ni_6$ specimens (RHEN-6Ni-1300SPS and RHEN-6Ni-1400SPS), all retained <150 nm grain sizes with >96% relative densities after five hours annealing at 1300 °C. They together represent some of the highest temperature stability achieved for dense bulk nanoalloys and ultrafine-grained metals.

**E. Discussion of Mechanisms**

The first main observation of this study is that addition of Ni (including Ni doping as well as Ni and Zr co-doping) can substantially improve the sintering of MoNbTaToW-based RHENs. Prior studies showed that Ni promoted activated sintering of W and Mo via the formation of liquid-like interfacial phases at GBs (that can be explained as segregation-induced GB premelting)[29-34] and this mechanism were also found to be responsible for sub-eutectic activated sintering of ceramics like $ZnO-Bi_2O_3$ [37] and $TiO_2-CuO$ [38]. In fact, it is well known that addition of Ni and other transition metals like Fe and Co can promote activated sintering of various refractory metals with proposed mechanisms based on segregation-enhanced GB diffusion (in general).[29-34,39-44] Unfortunately, the nanoscale grain sizes in our RHENs made it infeasible to prepare good enough transmission electron microscopy (TEM) specimens to directly characterize the GBs. Nonetheless, it is likely that a mechanism similar to that enables Ni activated sintering of W and Mo [29-34,39-44] is also responsible for Ni enhanced sintering of MoNbTaTiW-based RHENs. While Zr segregation is likely too refractory to promote the formation of liquid-like GBs (as Zr is not a good activated sintering aid for Mo and W), co-doping RHENs with Ni and Zr can in principle enhance GB disordering via coupling effects (that were recently discovered in HEAs via atomistic simulations and machine learning[45]). Future investigations, *e.g.*, TEM characterization of coarse-grained alloys of the same compositions that are equilibrated and well-quenched from the high sintering temperatures as well as thermodynamic and atomistic modeling (that are non-trivial and beyond the scope of this study), may be conducted to further test the hypothesis and the underlying mechanism of Ni enhanced sintering of MoNbTaTiW-based RHENs (that should occur via enhancing GB diffusion, regardless the actual GB structures).

Activated sintering via forming liquid-like GBs (or promoting GB diffusion in general)[29-34,39-44] would also likely promote grain growth. The second (perhaps even more interesting) observation of this study is that the addition of Ni, a well-known sintering aid for enabling activated sintering of W and Mo (that are also expected to promote grain growth),[29-34] in MoNbTaTiW-based RHENs can promote sintering without accelerating grain growth. Here, one possible mechanism is through the so-called "Type II HEGBs" (*i.e.*, HEGBs in HEAs),[27,28] following a hypothesis[27,28] that adding a small but saturated amount of one strong segregating element in a HEA can provide improved high-temperature stability against grain growth, where the effect is more significant for a larger number of principal components in the bulk phase.

The key concept of Type II HEGBs[28] can be illustrated using a simplified model as follows (where we expand a more brief derivation originally given in a recent Perspective article[28] and apply it to an idealized system to mimic our specific RHENs). Here, we adopt a multicomponent GB segregation model[26] generalized from the binary Wynblatt-Chatain model[36] to consider segregation in the two layers at the GB core of a general (large-angle) twist GB. To mimic MoNbTaTiW-Ni, let us consider a simplified, symmetric, ($N$+1)-component ideal solution, where 1, 2, …, and $N$ are (hypothetically) identical principal components ($M_i$) and ($N$+1) is a segregating component (S) and all binary $i$-($N$+1) systems ($i = 1, 2, ..., N$) have identical thermodynamic parameters. We assume, for simplicity, the ($N$+1)$^{th}$ component S is the only segregating component, so that a McLean-Langmuir type segregation equation applies:

$$\frac{X_{N+1}^{GB}}{X_{N+1}^{bulk}} = \frac{X_{1}^{GB}}{X_{1}^{bulk}} \exp\left(-\frac{\Delta g^{seg.}}{kT}\right) \approx \frac{X_{1}^{GB}}{X_{1}^{bulk}} \exp\left(-\frac{\Delta h^{seg.}}{kT}\right), \tag{1}$$

where $X_{i}^{bulk}$ and $X_{i}^{GB}$ are the bulk and GB compositions of the $i^{th}$ element, respectively, and $\Delta g^{seg.}$ ($\approx \Delta h^{seg.}$) is the free energy (enthalpy) change of the GB segregation of the ($N$+1)$^{th}$ component. Here, we adopt the sign convention that $\Delta h^{seg.} < 0$ for positive GB segregation. We further assume:

$$\begin{cases} X_{N+1}^{bulk} \ll 1 \\ X_{i}^{bulk} = X_{1}^{bulk} = \frac{1 - X_{N+1}^{bulk}}{N+1} \quad \text{(for all } i = 1, 2, ...N) \\ N \cdot X_{1}^{bulk} + X_{N+1}^{bulk} = 1 \end{cases} \tag{2}$$

Then, we can derive from Gibbs adsorption theory (based on the lattice-type multicomponent GB segregation model[26]) that GB energy is reduced with GB segregation, and further modified by the configurational entropy change at the GB, according to:

$$\begin{aligned}
\gamma_{\mathrm{GB}} &= \gamma_{\mathrm{GB}}^{(0)} + 2n_{\mathrm{PD}} \left[ X_{N+1}^{\mathrm{GB}} \Delta h^{\mathrm{seg.}} + kT \sum_{i=1}^{N+1} X_i^{\mathrm{GB}} \ln\left(\frac{X_i^{\mathrm{GB}}}{X_i^{\mathrm{bulk}}}\right) \right] \\
&= \gamma_{\mathrm{GB}}^{(0)} + 2n_{\mathrm{PD}} \left\{ X_{N+1}^{\mathrm{GB}} \left[ \Delta h^{\mathrm{seg.}} + kT \ln\left(\frac{X_{N+1}^{\mathrm{GB}}}{X_{N+1}^{\mathrm{bulk}}}\right) \right] + NkTX_1^{\mathrm{GB}} \ln\left(\frac{X_1^{\mathrm{GB}}}{X_1^{\mathrm{bulk}}}\right) \right\},
\end{aligned} \quad (3)$$

where $\gamma_{\mathrm{GB}}^{(0)}$ is the reference GB energy of an undoped GB ($X_{N+1}^{\mathrm{bulk}} = 0$), $n_{\mathrm{PD}}$ is the planar density of atoms (noting there are two segregating layers at the twist GB considered here), $k$ is the Boltzmann constant, and $T$ is temperature. Plugging in Eq. (1) and taking Eq. (2), Eq. (3) can be simplified to:

$$\begin{aligned}
\gamma_{\mathrm{GB}} &\approx \gamma_{\mathrm{GB}}^{(0)} + 2n_{\mathrm{PD}} \left\{ X_{N+1}^{\mathrm{GB}} \left[ \Delta h^{\mathrm{seg.}} + kT \ln\left(\frac{X_1^{\mathrm{GB}}}{X_1^{\mathrm{bulk}}}\right) - \Delta h^{\mathrm{seg.}} \right] + NkTX_1^{\mathrm{GB}} \ln\left(\frac{X_1^{\mathrm{GB}}}{X_1^{\mathrm{bulk}}}\right) \right\} \\
&= \gamma_{\mathrm{GB}}^{(0)} + 2n_{\mathrm{PD}}kT \left[ X_N^{\mathrm{GB}} \ln\left(\frac{X_1^{\mathrm{GB}}}{X_1^{\mathrm{bulk}}}\right) + NX_1^{\mathrm{GB}} \ln\left(\frac{X_1^{\mathrm{GB}}}{X_1^{\mathrm{bulk}}}\right) \right] \\
&= \gamma_{\mathrm{GB}}^{(0)} + 2nkT \ln\left(\frac{X_1^{\mathrm{GB}}}{X_1^{\mathrm{bulk}}}\right)
\end{aligned} \quad (4)$$

We further assume that all binary solvus lines are pinned by stoichiometric $(M_i)_1(S)_1$ precipitation compounds (assumed 1:1 stoichiometry for simplicity) following:

$$X_{i,\,\mathrm{binary\ solvus}}^{\mathrm{bulk}} \approx \exp(\Delta h^{\mathrm{sol./ppt.}} / kT), \quad (5)$$

where $\Delta h^{\mathrm{sol./ppt.}}$ ($= \Delta h_{i-(N+1)}^{\mathrm{sol./ppt.}}$, assumed to be identical for all $i = 1, 2, ..., N$ for simplicity) $< 0$ is the enthalpy of dissolving an atom of the $(N+1)^{\mathrm{th}}$ element from the $(M_i)_1(S)_1$ precipitate in the binary $i$-$(N+1)$ system. Thus, the maximum bulk solubility limit of the segregating $(N+1)^{\mathrm{th}}$ element in the $(N+1)$-component system (obtained at $X_i^{\mathrm{bulk}} = (1 - X_{N+1}^{\mathrm{bulk}})/N$ for all $i = 1, 2, ..., N$) is given by:

$$X_{N+1}^{\mathrm{bulk}} = N \cdot \exp\left(\frac{\Delta h^{\mathrm{sol./ppt.}}}{kT}\right) \quad (6)$$

Here, we assume that our doping level of the $(N+1)^{\mathrm{th}}$ element S is greater than $X_{N+1}^{\mathrm{bulk}}$ so that the fraction of segregating element S in grains is on the maximum multicomponent solvus line at equilibrium (*i.e.*, S-saturated grains in an equilibrium with $N$ precipitated phases). Comparing Eq. (5) and Eq. (6) suggests that a bulk high-entropy effect that enhances the maximum solubility of the segregating $(N+1)^{\mathrm{th}}$ element S, which increases with the number of the principal components ($N$), which consequently enables Type II HEGBs to stabilize GB segregation at high temperatures and reduce GB energy with increasing $N$, as elaborated below. Combining Eq. (6) with Eq. (1) (and utilizing Eq. (2) and $N \cdot X_1^{GB} + X_{N+1}^{GB} = 1$), we can obtain:

$$X_{N+1}^{GB} = \frac{N \exp\left(-\frac{\Delta h_N^{seg.-ppt.}}{kT}\right)}{1 + N \exp\left(-\frac{\Delta h_N^{seg.-ppt.}}{kT}\right)}, \tag{7}$$

where $\Delta h^{seg.-ppt.}$ ($\equiv \Delta h^{seg.} - \Delta h^{sol./ppt.}$) represents the enthalpy difference between the segregation and precipitation per atom. Eq. (7) suggests that the amount of GB segregation increases with the number of the principal components (as $N\uparrow$) due to bulk high-entropy effects shown in Eq. (6) that increases the solid solubility of the segregating $(N+1)^{th}$ element S.

Plugging Eq. (7) and $X_1^{GB} = (1 - X_{N+1}^{GB})/N$ into Eq. (4) and taking an approximation $X_1^{bulk} \approx 1/N$, we can derive an approximated analytical expression for an ideal solution at the dilute limit for a Type II HEGB in a HEA:

$$\gamma_{GB} \approx \gamma_{GB}^{(0)} - 2n_{PD}kT \ln\left[1 + N \cdot \exp\left(-\frac{\Delta h_N^{seg.-ppt.}}{kT}\right)\right]. \tag{8}$$

An empirical relationship exists: $\Delta h_N^{seg.} - \Delta h_N^{sol./ppt.} \approx -(0.10 \pm 0.06)$ eV/atom $\approx -(10 \pm 6)$ kJ/mol,[46] so that the term $\exp(-\Delta h_N^{seg.-ppt.}/kT)$ is greater than 1 (~2.15 at 1300 °C, if we takes the medium value of 0.1 eV/atom). Eq. (8) suggests more reduction in GB energy will reduce with increasing temperature ($\gamma_{GB}\downarrow$ as $T\uparrow$) and/or increasing the number of the principal components ($\gamma_{GB}\downarrow$ as $N\uparrow$).

Eq. (8) suggests more reduction in GB energy with more principal elements ($\gamma_{GB}\downarrow$ as $N\uparrow$), which will provide thermodynamic stabilization (reducing the grain growth driving force). Eq. (7) further suggests more GB segregation with more principal elements ($X_{N+1}^{GB}\uparrow$ as $N\uparrow$), which will provide more kinetic stabilization via solute drag. Moreover, these bulk high-entropy effects in Type II HEGBs are more effective (to reduce GB energy $\gamma_{GB}$ and increase GB segregation $X_{N+1}^{GB}$) at high temperatures (because $\exp(-\Delta h_N^{seg.-ppt.}/kT)$ is a positive factor in both Eq. (7) and Eq. (8)). In other words, both the above hypothesized thermodynamic and kinetic stabilization mechanisms can be more effective with more principal elements (as $N\uparrow$) at high temperatures.

Consequently, this Type II HEGB effects can serve as one possible mechanism in stabilizing Ni-doped MoNbTaTiW RHENs against rapid grain growth (where $N = 5$ and Ni is the 6$^{th}$ segregating element, which can offer more stabilization effects than that in binary W-Ni and Mo-Ni alloys with $N = 1$, based on the above simplified model). We should note that the real RHENs are more complex than the above simplified ideal model. Yet, the simplified and symmetric ideal solution model allows us to derive an analytical solution to illustrate a key mechanism to inhibit grain growth at high temperatures in HEAs with one strong segregating element via Type II HEGBs. This mechanism offers one possible explanation of our observation of exceptional high-temperature stability in Ni-doped RHENs ($Mo_{18.8}Nb_{18.8}Ta_{18.8}Ti_{18.8}W_{18.8}Ni_6$ or RHEN-6Ni). This stabilization effects via Type II HEGBs can be further enhanced with co-doping (*e.g.*, in RHEN-6Ni-5Zr and RHEN-3Ni-2.5Zr). We also recognize that secondary phase precipitates (from either the Mo-Nb-Ta-Ti-W-Ni-Zr systems or native oxides or impurities from ball milling) can also suppress grain growth via Zener pinning. Further investigations may quantify different stabilization mechanisms (albeit highly challenging given the complexity of real RHENs). We recognize the real RHENs are much more complex than the simplified model; nonetheless, an approximated analytical solution can be obtained from the simplified to illustrate one of (but perhaps not the only) the possible mechanisms of high-temperature stability against rapid grain growth in Ni-doped RHENs (observed in this study), as well as potentially many other high-entropy nanoalloys to be investigated in future studies.

## IV. CONCLUSIONS

In summary, this study has simultaneously achieved exceptional high-temperature stability and improved sinterability to fabricate a novel class of MoNbTaTiW-based refractory high-entropy nanoalloys (RHENs). The effects of adding Ni and Zr in sintering and grain growth have been carefully examined. The fabricated bulk specimens (1 cm in diameter and ~3 mm in thickness) have achieved 93-96% relative densities with 50-100 nm grain sizes for three different compositions. For example, a $Mo_{17.8}Nb_{17.8}Ta_{17.8}Ti_{17.8}W_{17.8}Ni_6Zr_5$ fabricated by SPS at 1300 °C attained ~96% relative density with ~55 nm mean grain size. We further demonstrated that four different RHENs retained <150 nm grain sizes with >96% relative densities after five hours annealing at 1300 °C, which represent some of the highest temperature stability that has been achieved for nanoalloys and ultrafine-grained metals.

Notably, the addition of Ni, a well-known sintering aid for activated sintering of refractory metals such as W and Mo (that also promote grain growth), in high-entropy MoNbTaTiW can promote sintering while maintaining high-temperature stability against rapid grain growth. This latter unusual observation can be explained by hypothesized effects of high-entropy grain boundaries (Type II HEGBs in HEAs), for which we further elaborated the relevant theory with an idealized model in this work. We recognize other stabilization mechanisms, such as Zener pinning, can also be present in these complex RHENs.

## ACKNOWLEDGEMENTS

We gratefully acknowledge the support by the U.S. Army Research Office (Grant No. W911NF2210071, managed by Dr. Michael P. Bakas, in the Synthesis & Processing program).

## AUTHOR DECLARATIONS

**Conflict of Interest** The authors have no conflicts to disclose.

## AUTHOR CONTRIBUTIONS

M.Q. conducted most experiments and some data analysis. S.S. helped with some characterization and analysis and drafted an initial manuscript. J.L. conceived the idea, supervised the study, and rewrote the manuscript.

## DATA AVAILABILITY

The data that support the findings of this study are available within the article and its supplementary material.

## SUPPLEMENTARY MATERIAL

See supplementary Tables S1-S2 and Figs. S1-S10 for detailed sintering and grain growth annealing experiments data, selected XRD patterns, additional SEM images of fractured surfaces before and after annealing, and relevant binary phase diagrams.

**Table 1.** Summarized results of selected as-sintered RHENs, including compositions, SPS conditions, measured relative densities, and measured average grain sizes (± 1 standard deviation). Complete data for all 26 as-sintered RHENs made in this study are documented in Supplementary Table S1.

| Composition | SPS Conditions | | Relative Density | Grain Size |
|---|---|---|---|---|
| RHEN-6Ni (18.8Mo-18.8Ti-18.8Ta-18.8Nb-18.8W-6Ni) | 1300 °C × 5 min | Specimen #1 | 93.1% | 74 ± 19 nm |
| | | Specimen #2 | 93.6% | 78 ± 27 nm |
| | 1400 °C × 5 min | Specimen #1 | 94.9% | 103 ± 35 nm |
| | | Specimen #2 | 94.7% | 109 ± 40 nm |
| RHEA-6Ni-5Zr (17.8Mo-17.8Ti-17.8Ta-17.8Nb-17.8W-6Ni-5Zr) | 1300 °C × 5 min | | 96.0% | 55 ± 14 nm |

**Table 2.** Summarized results of the three selected RHENs after annealing at 1300 °C for 5 h, including measured relative densities, average grain sizes, and Rockwell hardness (HRC). More data for 28 annealed specimens in seven series of RHENs (*i.e.*, 21 annealed specimens, along with seven as-sintered RHENs before the annealing as the references) are documented in Supplementary Table S2.

| Composition and Processing History | Relative Density (%) | SEM Grain Size (nm) | Hardness (HRC) |
|---|---|---|---|
| RHEN-6Ni<br>1300 °C × 5 min SPS<br>+ 1300 °C × 5 h Annealing | 97.0 | 133 ± 28 | 64 ± 3 |
| RHEA-6Ni<br>1400 °C × 5 min SPS<br>+ 1300 °C × 5 h Annealing | 97.2 | 124 ± 10 | 65 ± 2 |
| RHEA-6Ni-5Zr<br>1300 °C × 5 min SPS<br>+ 1300 °C × 5 h Annealing | 96.3 | 133 ± 26 | 62 ± 4 |

**Table 3.** Summary of relevant thermodynamic data of $X$-Ni and $X$-Zr ($X$ = W, Mo, Ta, Nb, or Ti) binary alloys, including the solubilities at 1300°C, the lowest eutectic temperatures, and solidus temperatures for given compositions (taken from binary phase diagrams shown in Supplementary Fig. S9 and Fig. S10 from ASM Alloy Phase Diagram Database$^{TM}$) and the segregation enthalpy $\Delta H_{seg}$ values calculated from a Miedema-based model by Murdoch and Schuh.[35] Both Ni and Zr prefer to segregate at GBs in $X$ ($X$ = W, Mo, Ta, Nb, or Ti) in all ten binary systems. Here, we adopt a sign convention where a negative $\Delta H_{seg}$ value suggests positive GB segregation or GB enrichment[36] (differing from the sign convention used in the original Ref. 35 where a positive $H_{seg}$ infers GB segregation).

| $X$ | Solubility at 1300°C (at%) | | Lowest Eutectic Temperature $T_E$ (°C) | | Solidus Temperature $T_S$ (°C) | | Segregation Enthalpy $\Delta H_{seg}$ (J/mol) | |
|---|---|---|---|---|---|---|---|---|
| | Ni in $X$ | Zr in $X$ | $X$-Ni | $X$-Zr | $X$-6Ni | $X$-5Zr | $X$-Ni | $X$-Zr |
| W | 0.2 | 1.1 | 1500 | 1740 | 1500 | 2160 | -78276 | -58933 |
| Mo | 1.3 | 7 | 1312 | 1576 | 1360 | 1915 | -46901 | -47254 |
| Ta | 2.9 | 9.2 | 1366 | 1885 | 2700 | 2420 | -79917 | -30822 |
| Nb | 7 | 100 | 1158 | 1730 | 1278 | 1810 | -42351 | -19478 |
| Ti | 4.2 | 100 | 952 | 1554 | 1200 | 1800 | -24897 | -1280 |
| **Rule-of-Mixture (RoM) Average Value:** | | | | | 1608 | 2021 | -54468 | -31553 |

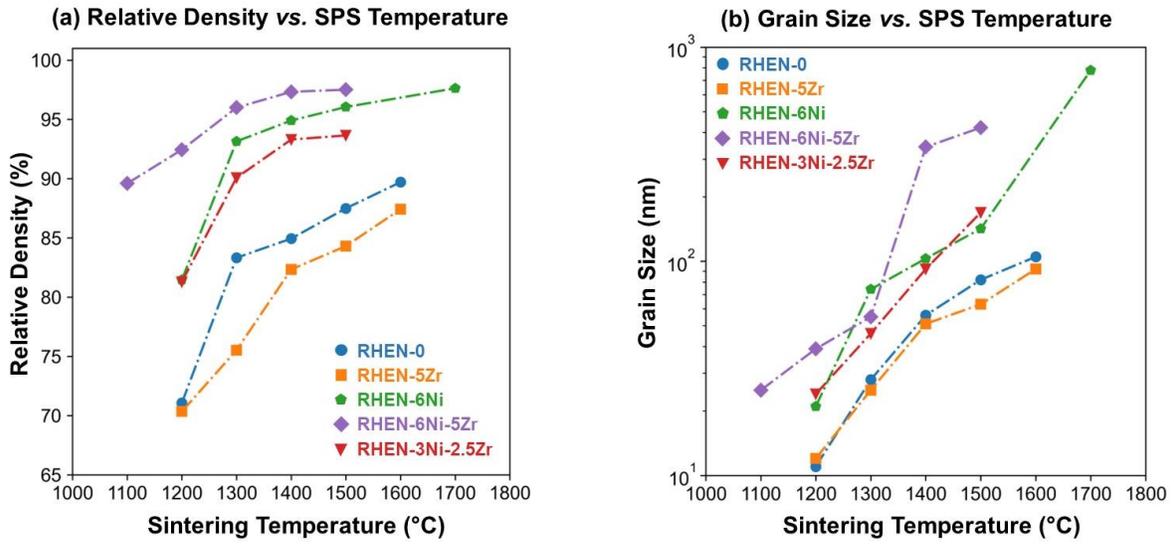

**FIG 1.** Measured **(a)** relative density and **(b)** grain size *vs.* SPS temperature curves for five series of REHNs after SPS. All specimens were hold isothermally at the sintering (SPS) temperatures for 5 min under 50 MPa of pressure.

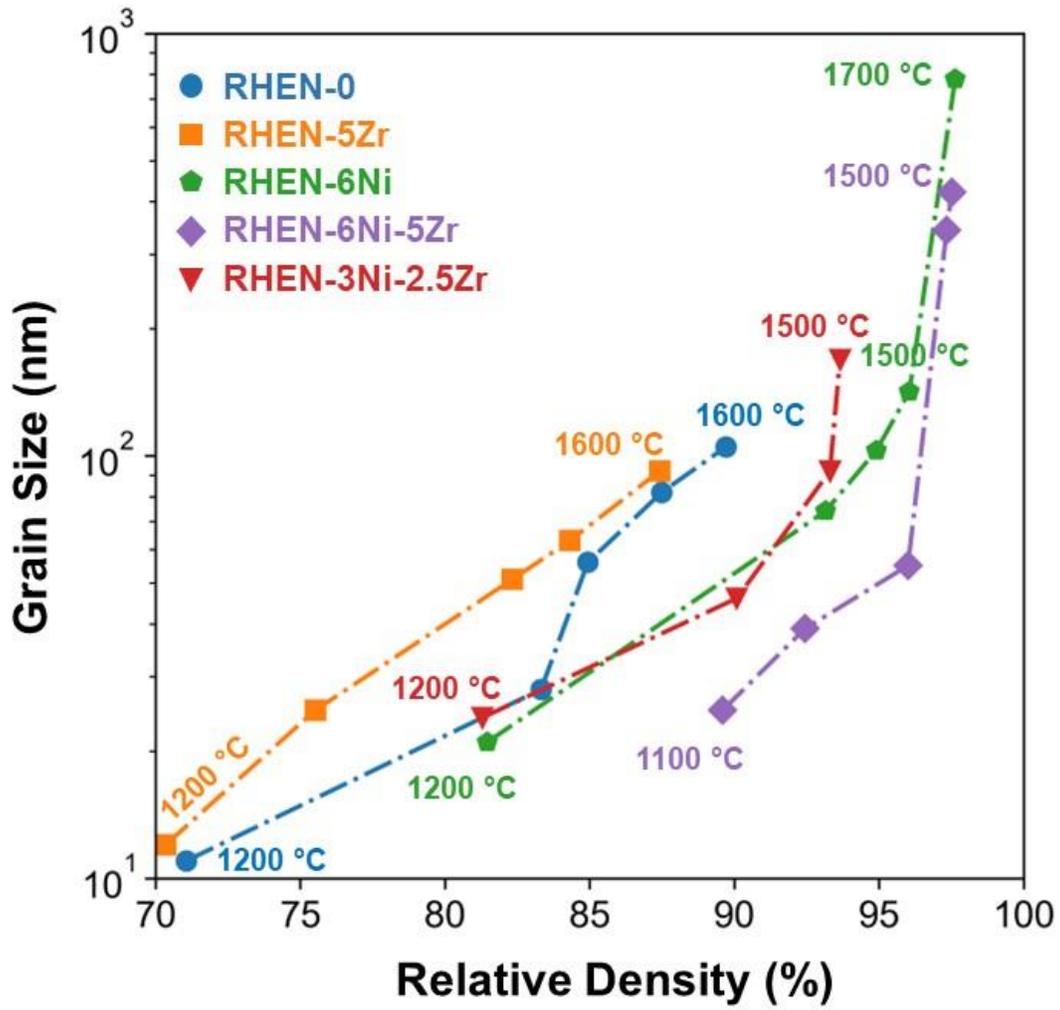

**FIG. 2.** Grain size *vs.* relative density curves for five series of REHNs after SPS.

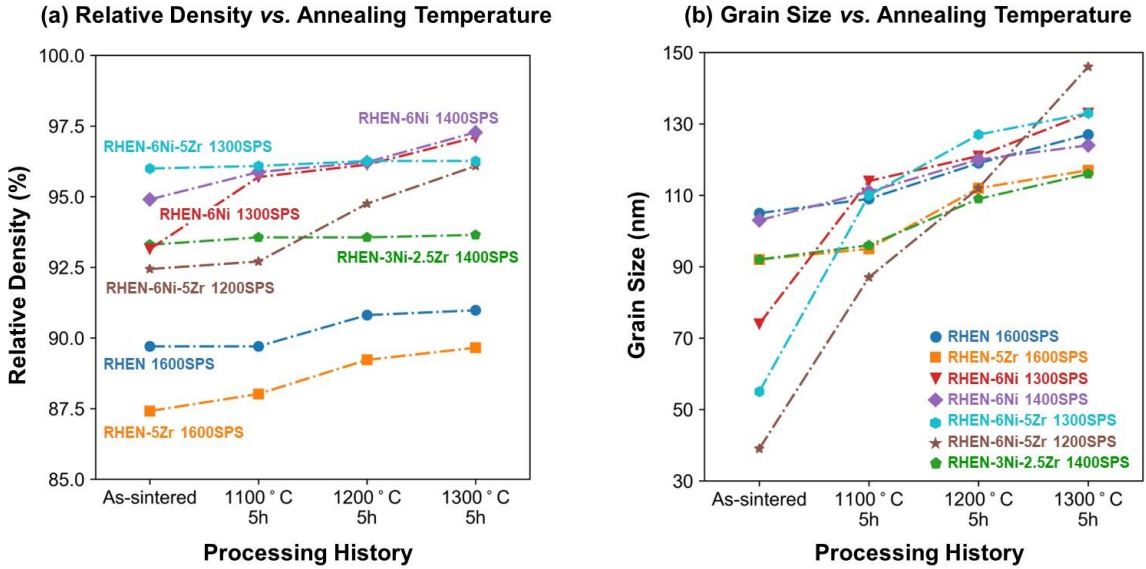

**FIG 3.** Measured **(a)** relative density and **(b)** grain size *vs.* annealing temperature curves for the grain growth experiments, where seven series of RHENs were annealed at 1100 °C, 1200 °C, and 1300 °C, respectively, isothermally for 5 hours. The corresponding data from the seven as-sintered RHENs before annealing are also shown as references.

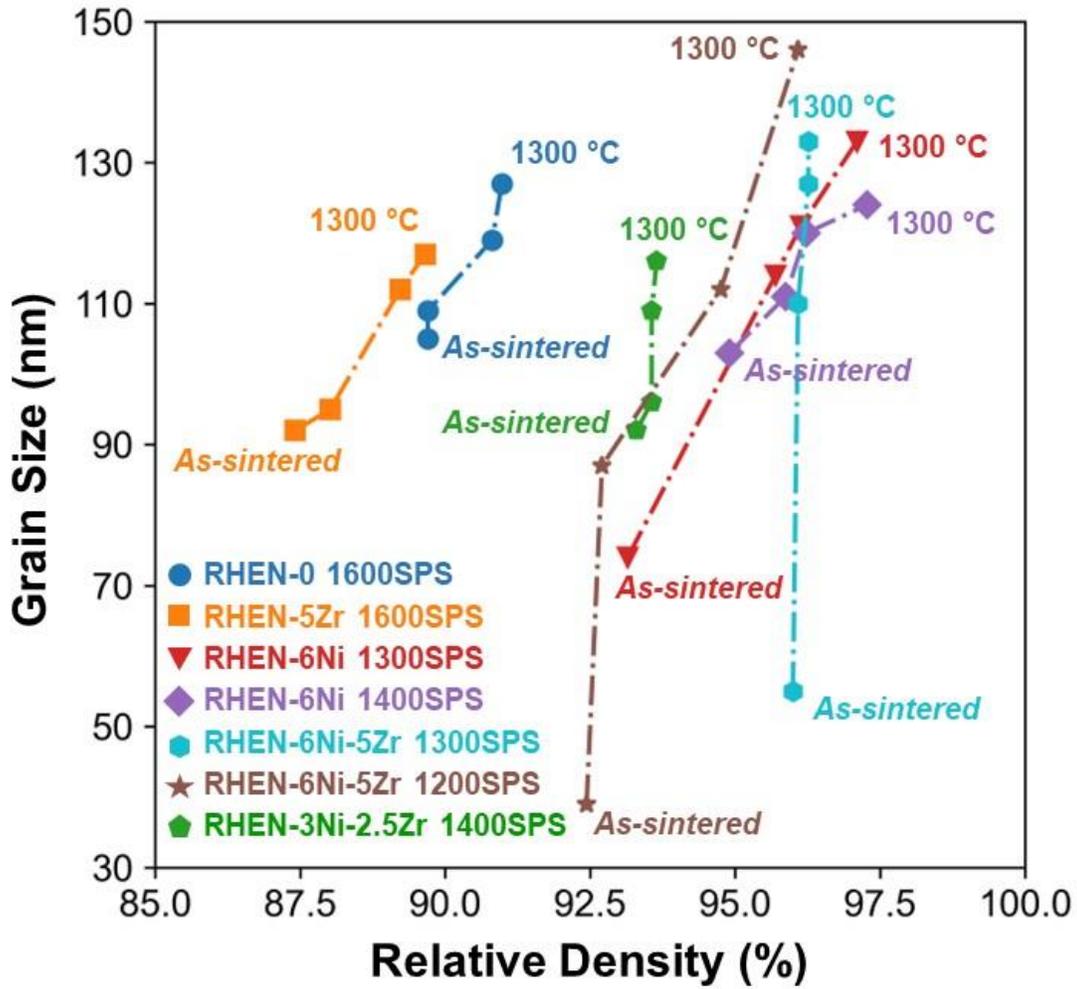

**FIG. 4.** Grain size *vs.* relative density curves for seven series of as-sintered RHENs and specimens after annealing at 1100 °C, 1200 °C, and 1300 °C, respectively.

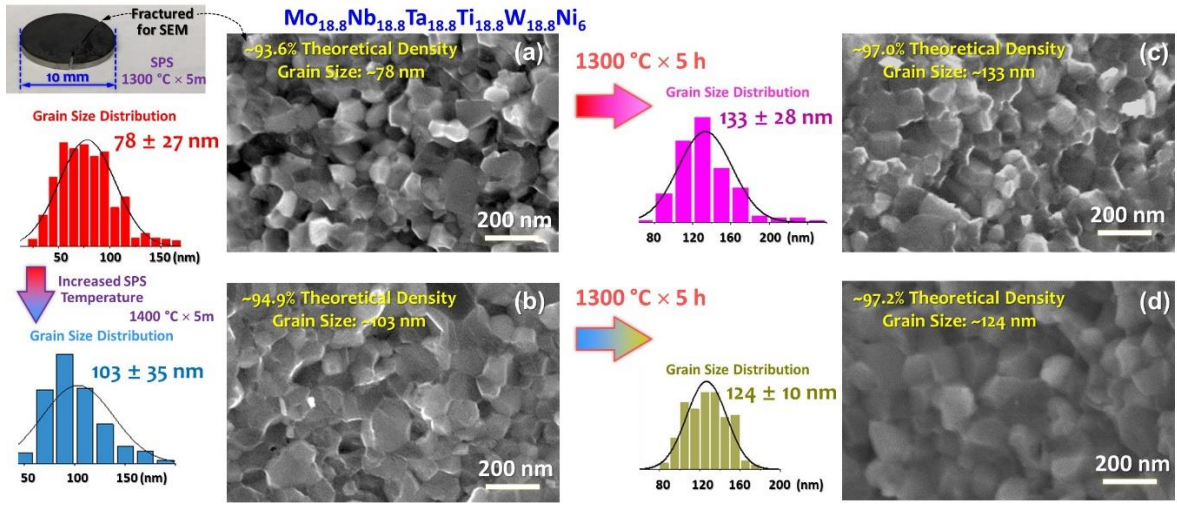

**FIG 5.** Grain size distributions and SEM micrographs of as-sintered RHEN-6Ni specimens sintered at **(a)** 1300 °C (~93.6%, ~78 nm grain size) and **(b)** 1400 °C (~94.9% relative density, ~103 nm grain size), respectively, and **(c, d)** after annealing at 1300 °C for 5 hours with increased density (both to ~97% relative densities) and limited grain growth (both <150 nm in grain sizes). See Supplementary Figs. S3-S6 for additional specimens and views of larger regions of as-sintered and annealed RHEN-6Ni specimens.

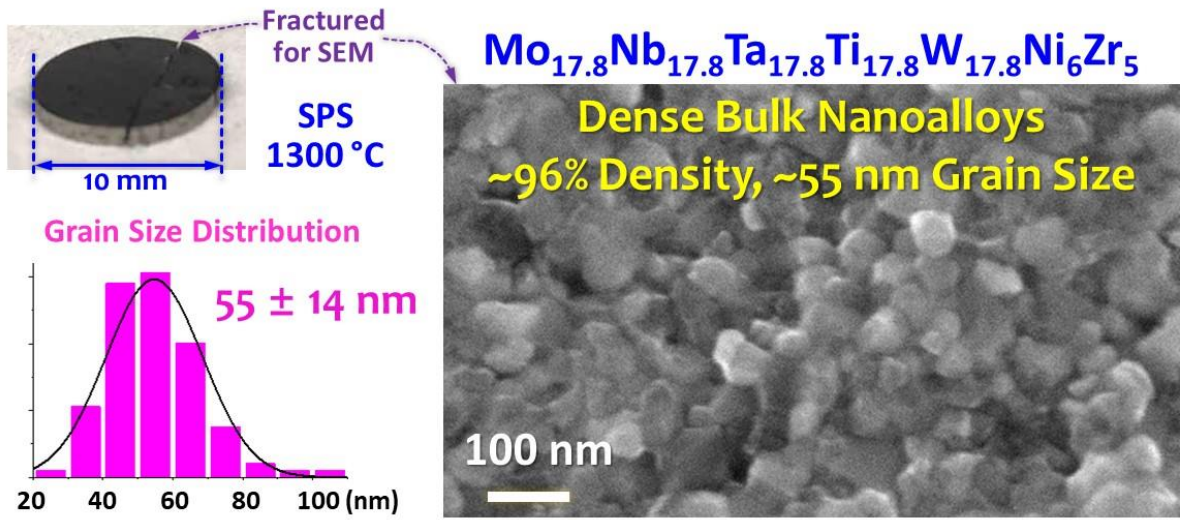

**FIG. 6.** Grain size distribution and SEM micrograph of a RHEN-6Ni-5Zr specimen sintered at 1300 °C, showing a high relative density of ~96% and a small mean grain size of ~55 nm. SEM micrograph of a large region of this specimen is shown in Supplementary Fig. S7.

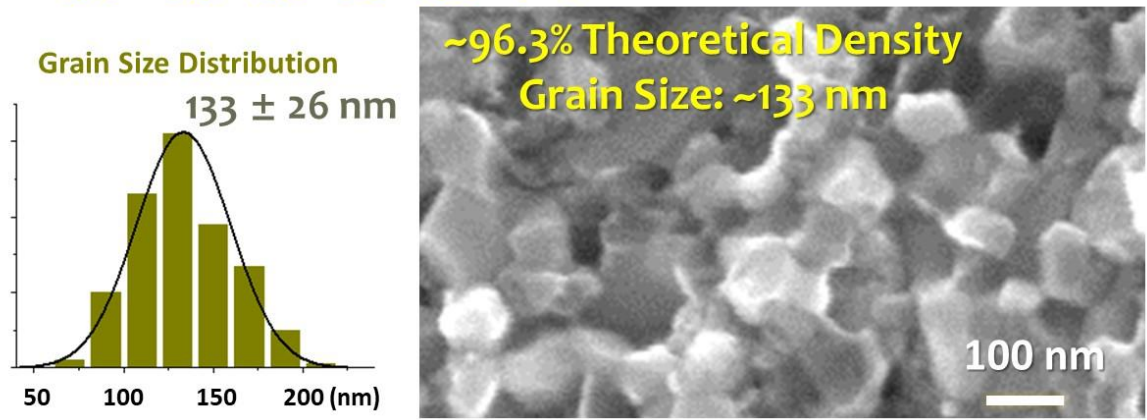

**FIG. 7.** Grain size distribution and SEM micrograph of a RHEN-6Ni-5Zr-1300SPS specimen after annealing at 1300 °C for 5 hours, showing ~96.3% relative density and a mean grain size of ~133 nm. SEM micrograph of a large region of this specimen is shown in Supplementary Fig. S8.